\documentclass[12pt]{iopart}
\usepackage{epsfig}

\begin{document}

\title[Heavy Flavours]{
Heavy Flavours in Collider Experiments
}

\author{R D St.Denis\dag}

\address{\dag\ University of Glasgow, Department of Physics and Astronomy,
Kelvin Building, Glasgow G12 8QQ, UK}

\begin{abstract}
Current issues in the studies of Heavy Flavours in colliders are described
with particular emphasis on experiments in which the UK is involved.  Results
on charm production at HERA are examined and
compared to those at the Tevatron.  $B$ production rates at the Tevatron as well 
as the status of $B$ lifetimes and mixing in the
LEP collaborations and at the Tevatron are highlighted. The measurement of
$\sin{2\beta}$ from CDF is described as well as the most recent 
results on top physics at the Tevatron.

\end{abstract}

\pacs{00.00, 20.00, 42.10}

\submitted to \jpg for publication.

\maketitle

\section{Introduction}

In this paper the status of heavy flavours, charm, bottom and top,
is reviewed.  For the charm quark, emphasis is placed on measurements of the
production rates and the differential cross sections at HERA and the Tevatron
and of the extraction of the gluon structure function from the proton at HERA
using charm.  The large kinematic range at HERA leads to a number of models
with limits on applicability.  These models cannot be directly applied to the
Tevatron and the work to understand charm production is not complete.
For bottom physics, the current status of the lifetime ratios
between charged and neutral mesons is examined in the context of 
 the spectator model.  The production cross section
at HERA and the Tevatron is also studied revealing an area that needs further
study.  Experimental methods for doing a $B$ mixing analysis
at hadron machines are explored and the results of a variety of $B$ mixing
studies presented.  This leads to a description of the measurement of the CP
violation parameter, $\sin{2\beta}$, at CDF as an indication of the propects
for measurements of this value at hadron machines.  A brief summary of the
current knowledge and world sample of top quarks closes the report.

\section{Charm at Colliders}
\begin{figure}[htb]
\begin{center}
  \includegraphics[width=10cm]{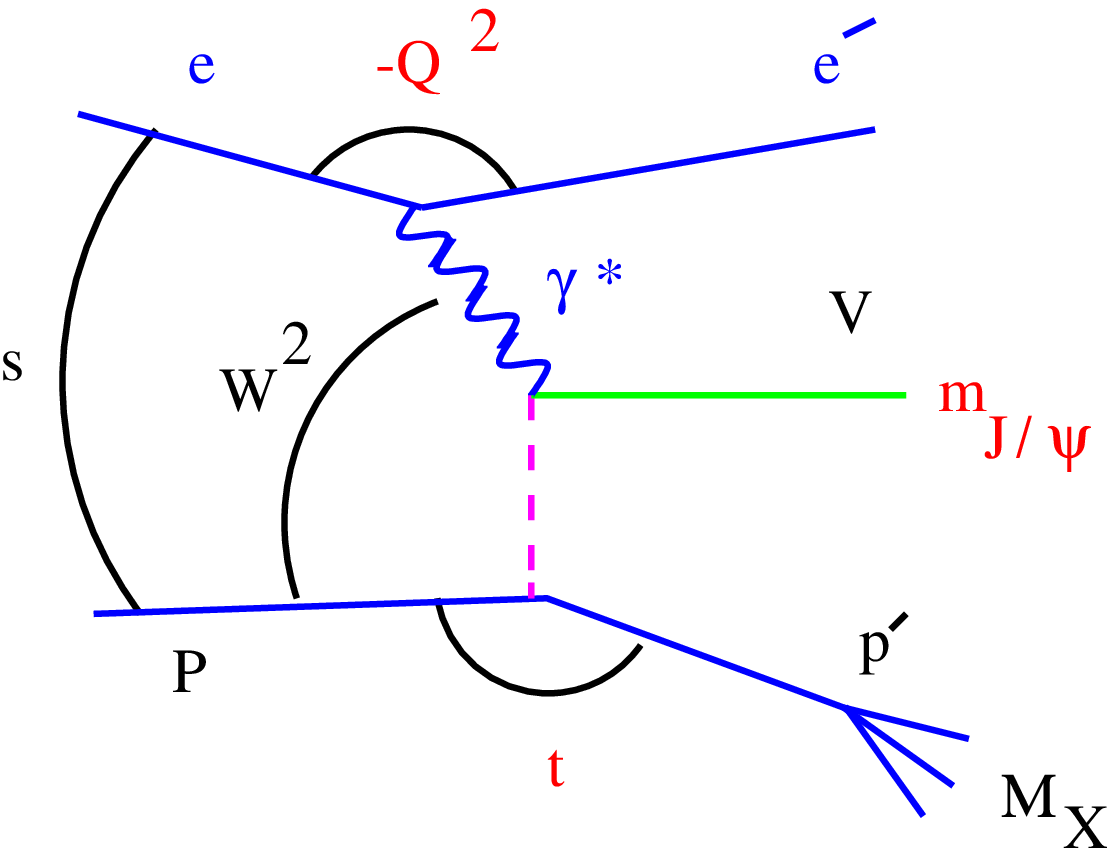}
  \includegraphics[width=15cm]{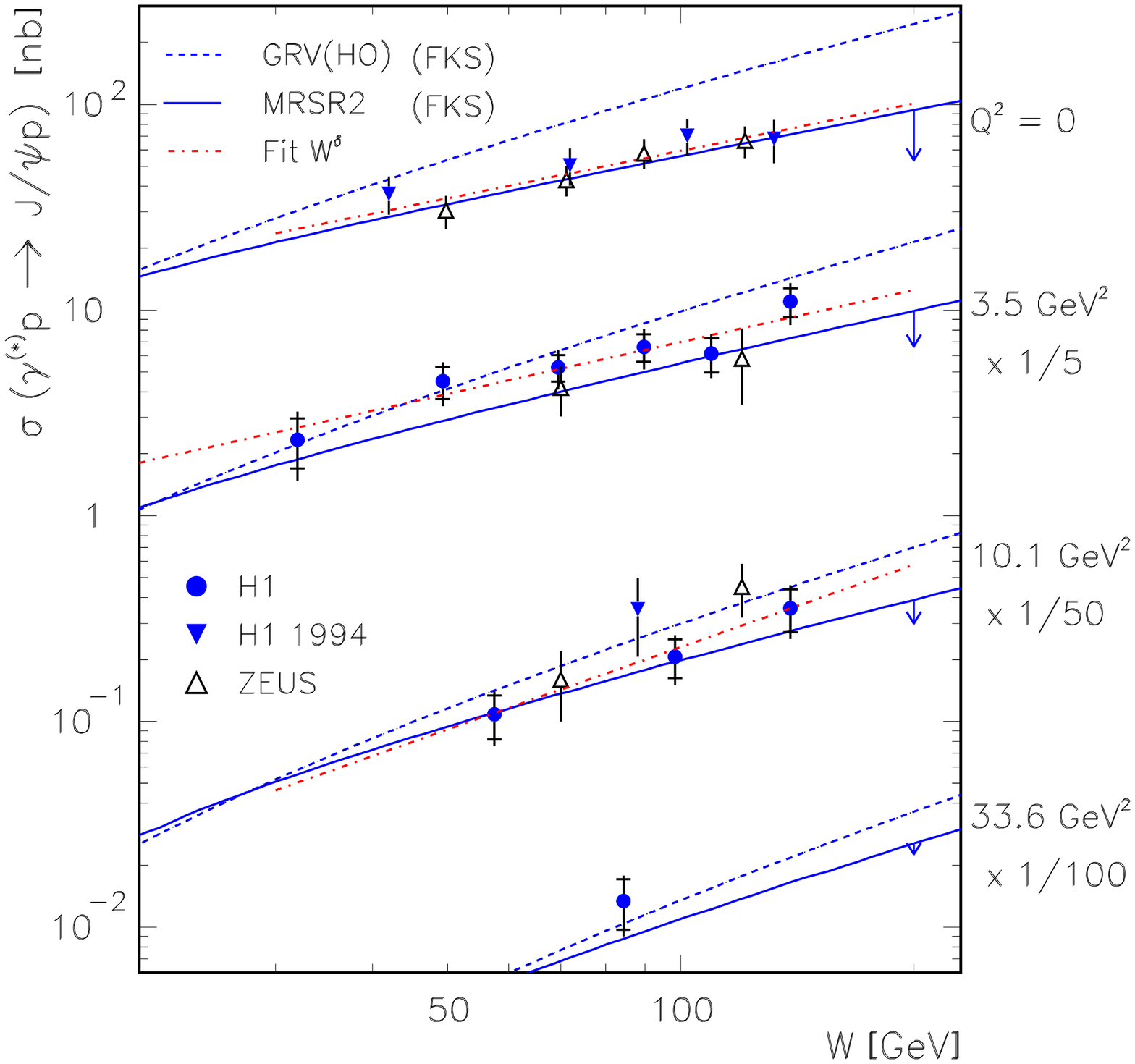}
  \caption{$J/\psi$ production kinematics at HERA and the
$W$ dependence of the differential cross section 
for $J/\psi$ production measured at HERA
 for a variety of $Q^2$ compared QCD.
}
\end{center}
\label{fig:disandwdepend}
\end{figure}

Figure~\ref{fig:disandwdepend} illustrates the relevent kinematic variables 
in vector boson, V,  production  at HERA for V=$J/\psi$:
{\it s}, the electron-proton center of mass energy,
{\it Q}, the photon virtuality,
{\it t}, the four momentum transfer to proton,
$M_x$, the mass of dissociated proton remnants,
{\it W}, the $\gamma$-proton CM energy.
By choosing various ranges in the kinematic variables, experimental
observations can be matched to theoretical predictions for the processes
expected to dominate the dynamics in the chosen region.
Various descriptions of charm production 
and the relationship between theoretical description
and experimental signature
are given in Table~\ref{tab:heramess} where the variable
{\it z}, the fraction of $\gamma$ momentum given to {\it J}/$\psi$, has been introduced. 
In each case a $J/\psi$ or a $D^*$ is observed.  The first column of the
table gives an additional experimental constraint.  The second column is the
conventional description of the process with the third column indicating
constraints on the kinematics.
\begin{center}
\begin{table}[hbt]
\caption{Guide to the HERA experimental description, common process
  name and
  kinematic cuts.}
\begin{indented}
\item[]\begin{tabular}{@{}lll}
\br
Experimental        & Common Name              & Kinematic    \\
Constraint          &                          & Restrictions \\
\mr
no electron         &photoproduction           &$Q^2=0$\\
an electron         &DIS                       &$Q^2>0$\\
other activity      &inelastic or 
 &$z<1$;$M_x>0$\\
                    &resolved photoproduction  &             \\
no other activity   &elastic                   &$z=1$;$M_x=0$\\
no forward activity &diffractive               &rapidity gaps\\
\br
\end{tabular}
\end{indented}
\label{tab:heramess}
\end{table}
\end{center}

Central to the description 
of charm production is the scale of the processes under scrutiny.
One suggested~\cite{ref:gluladderandscale} choice for the scale is 
$\mu^2 = \left[Q^2 + M_{J/\psi}^2 + \left|t\right|\right]/4$.
  This implies that for
applications of perturbative QCD, the minimum scale is the mass of the 
$J/\psi$.  
There is heated debate over whether perturbative QCD computations of
one or two gluon ladders can describe the 
data~\cite{ref:gluladderandscale,ref:restladder1,
ref:restladder2,ref:restladder3,ref:restladder4} or whether modified
models of soft Pomeron exchange~\cite{ref:pom} are more satisfactory. 
 One example where the perturbative QCD models describe the data
is illustrated
in Figure~\ref{fig:disandwdepend}~\cite{ref:wdep}. 
The data are also shown to have a $W$ dependence
such that the cross section goes as $W^{\delta}$ with $\delta \simeq 1$.
The Vector Dominance Model and 
Regge theory predict a smaller slope of $\delta=0.2$ and therefore do not
provide a good description of the data.

\begin{figure}[htb]
\begin{center}
\includegraphics[width=18cm]{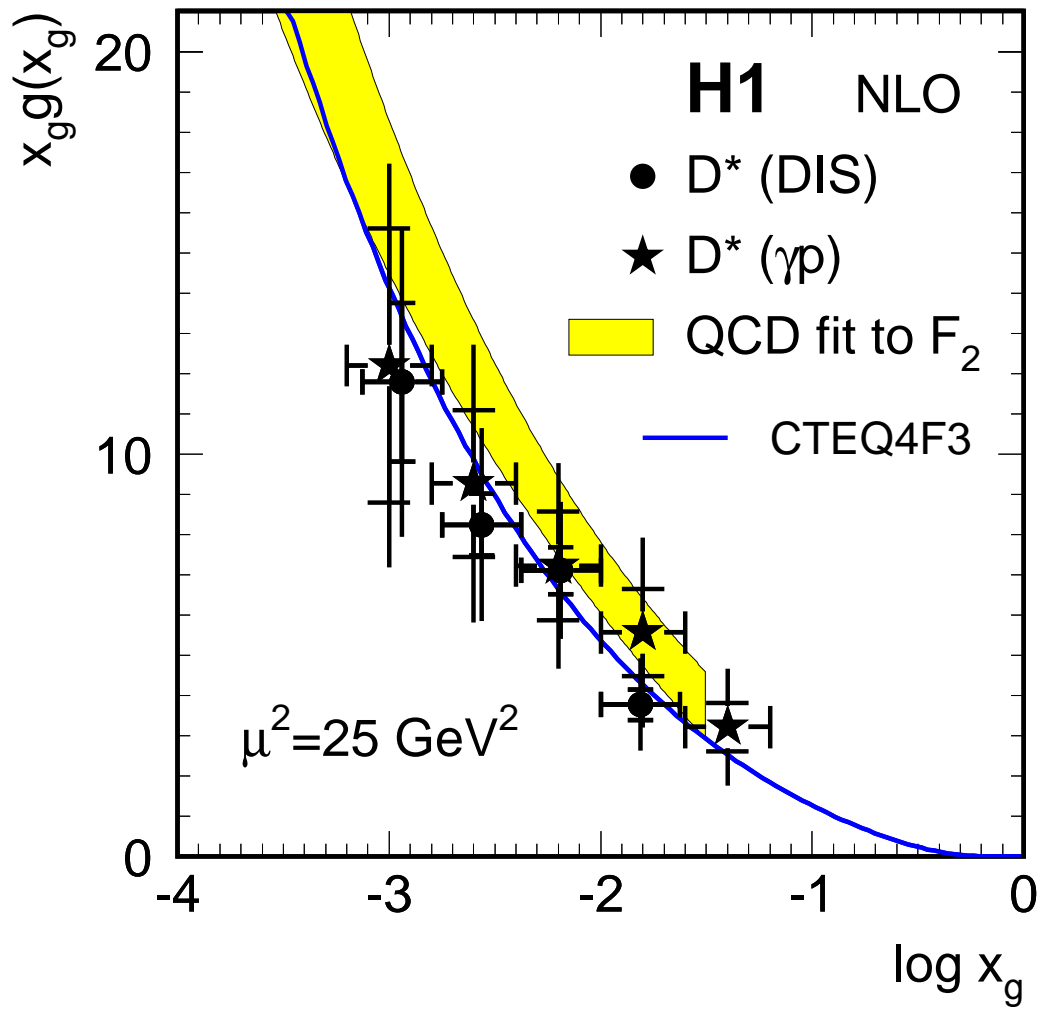}

\caption{Gluon Structure function: comparison of that derived from deep
  inelastic scattering of electrons (shaded band) and that derived from
   measurement of open charm.}\end{center}
\label{fig:gx}
\end{figure}

Further success in the description of the data can be found in the
measurement of the gluon structure function using open charm production and
comparing it to the measurement of 
$F_2(x,Q^2)$~\cite{ref:charmZeus,ref:charmH1}.  This is illustrated for the
H1  measurement in
Figure~\ref{fig:gx}. 
 The rapid rise of the gluon structure function at low $x$
is a prediction of its evolution in the context of the DGLAP 
equations~\cite{ref:DGLAP}.

However, application of perturbative QCD to the Tevatron Collider data does
not succeed~\cite{ref:CDFjpsi,ref:D0jpsi}. 
A colour singlet model~\cite{ref:colsing},
like that used successfully in HERA, fails.
The disagreement is
as large as a factor of 50.  Better agreement can be found if the CDF data
are fit with the colour octet model~\cite{ref:octet1,ref:octet2};
 however
the fits of this model fail badly when applied to
the differential cross sections measured in a variety of kinematic
variables at HERA~\cite{ref:wdep}. 

While it is tempting to dismiss the scale set by the charm mass as too small
and the discrepancies as being no surprise, doing so would ignore the results
where there is success in the predictions.  Assembling the successes and
failures and understanding the limits of the use of perturbative QCD and its
matching with soft phenomenology remains an unsolved problem.

\section{Bottom at Colliders}

The production of $b$ quarks in hadron machines differs from the intuitively 
appealing $Z$ decay mechanism resulting in two 45 GeV jets having $B$ mesons
carrying 80\% of the jet energy.
In hadron machines the trigger biases the kinematics.
 One trigger used by CDF 
is a lepton having high transverse momentum relative to the 
proton-antiproton collision.  The lepton can be 
 either an electron or muon from a semileptonic decay.  
The $B$ mesons typically have a transverse momentum of 
$P^B_t= 20-25$ GeV/c. 

 Another very
important trigger is a dimuon trigger with a cut on the invariant mass of
the muons. This $J/\psi$ trigger, so called because the invariant mass of the
 muons is required to be within a broad range around the $J/\psi$ mass,
 will yield $B$ mesons having a $P^B_t = 10$  GeV/c.  
Alternatively the $J/\psi$ mass may be
excluded in order to get a cascade of sequential decays: $b\rightarrow \mu +
 c $ followed by $c \rightarrow \mu + X$.  The $B$ mesons or baryons carry 70\% of the $b$ quark
 energy but the particles may or may not be recognizable as jets.  The hadron
 colliders then rely on lepton tagging, displaced vertices and reconstructed
 decay products to perform the measurements. 
 In this regard, once the trigger provides a $b$-enriched sample, 
the tools for finding $b$'s are similar in hadron and lepton colliders.
Narrow resonances are particularly useful because they reduce the
combinatoric possibilities.
From such samples, a variety of measurements have been performed and in the
following sections a few have been selected for exposition.

\subsection{$B$ Lifetimes}
One measurement of particular significance is that of the various $B$ meson and
baryon lifetimes~\cite{ref:lifewg}. 
The ratio of $b$ lifetimes is important because it probes the extent to which
the simple spectator model 
is
valid and the degree to which the other diagrams including $W$ exchange and
final state interference
play a role.  
The measured value of the charged to neutral meson
lifetime ratio is $1.07 \pm 0.02$, well within 
the theoretical prediction of 1.0-1.1.
Thus, to a precision of a few per-cent, the spectator model is quite good
 and the more subtle effects are only beginning to be seen.
There is some discrepancy for the baryon lifetimes; however, their errors
 remain statistically dominated and rigorous theoretical tests await
the larger data samples anticipated at the next run of the Tevatron.

\subsection{$B$ Production Cross Section}

\begin{figure}[htb]
   \begin{center}
\includegraphics[bb=37 122 514 660,clip,width=10cm]{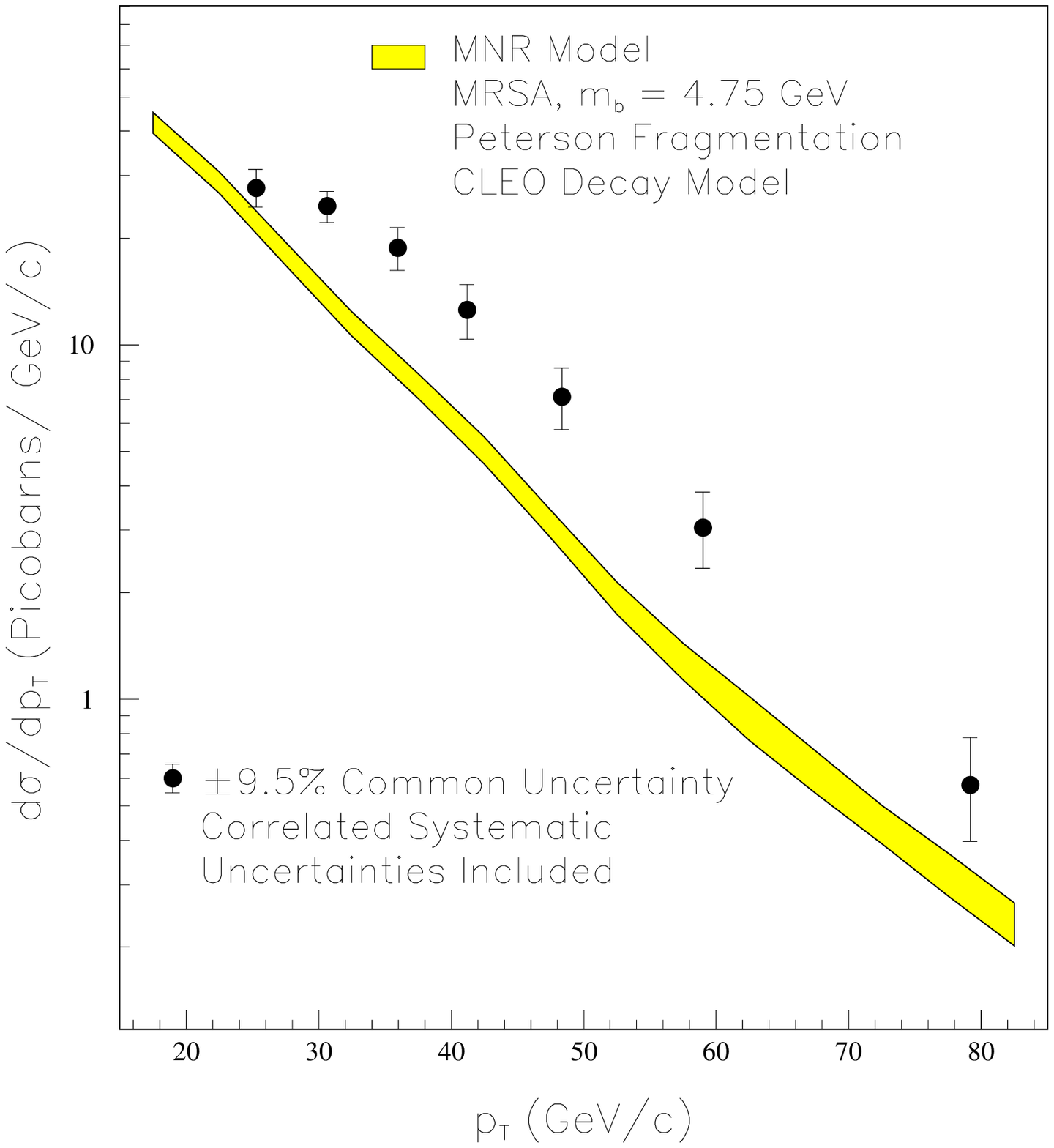}
\includegraphics[bb=37 122 514 660,clip,width=10cm]{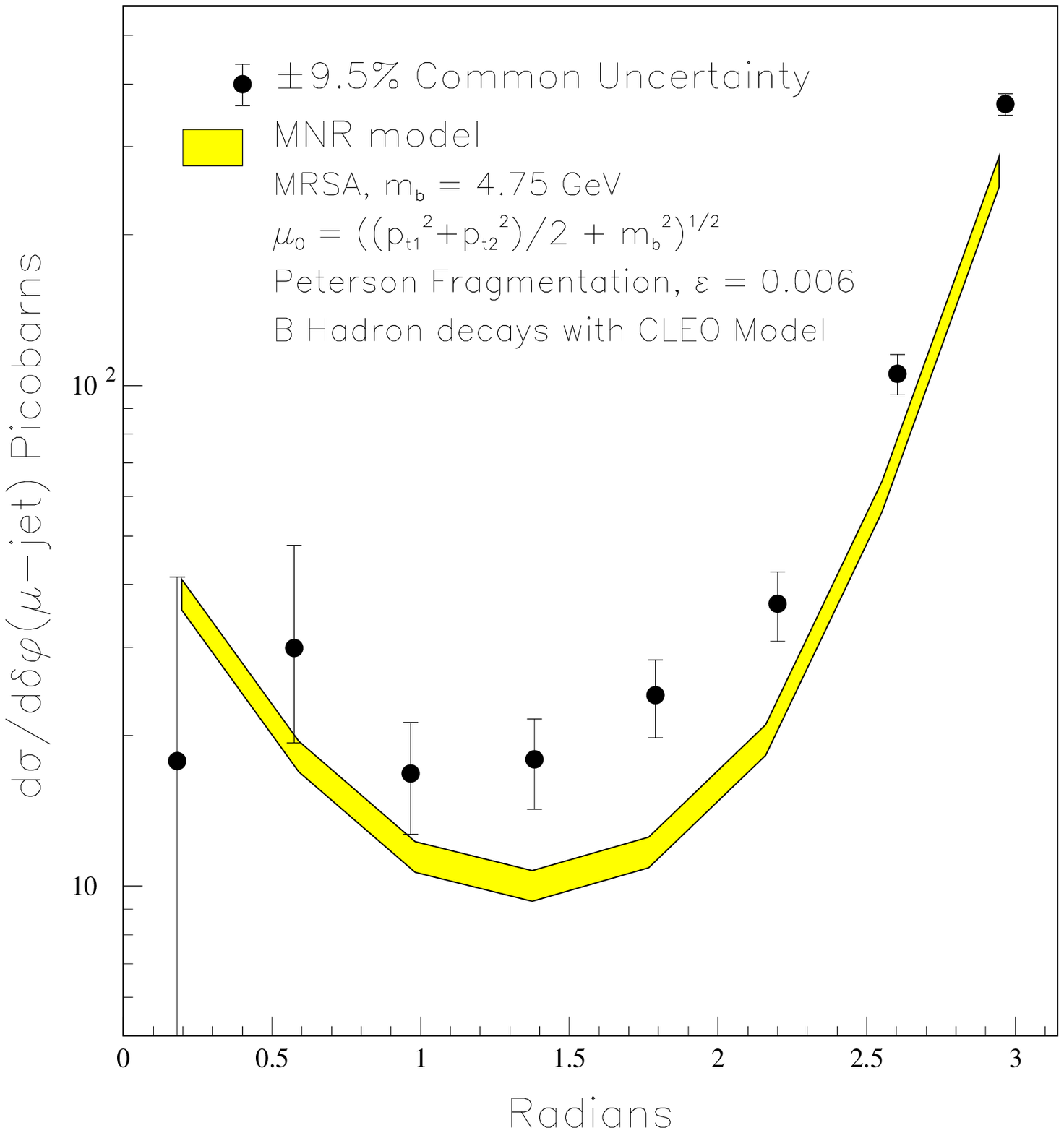}
\end{center}
\caption{Measurement of the the differential $b$ cross section in CDF and the
Azimuthal separations of the muon from semileptonic decay of a b
and the jet from the $\bar{b}$ at CDF.}
\label{fig:bopencrossect}
\end{figure}

As with the charm quark, theoretical computation of inclusive 
bottom production cross sections is not well understood.
For the production of open $b$,
as illustrated in Figure~\ref{fig:bopencrossect}, the situation is as
follows. The D0 collaboration~\cite{ref:D0Bxsect} finds that the shape is well described 
 but the cross section is
a factor of two to three higher than the theory predicts.
The H1 collaboration~\cite{ref:H1Bxsect} used inclusive muon production 
and reported a production cross section about a factor of five above theoretical prediction.
  This analysis has
subsequently been revised~\cite{ref:H1NewBxsect} and is now only 
two times the prediction.
 The CDF collaboration finds that the cross sections for
 $B$ $\rightarrow J/\psi K^+$,$K^{*0}$~\cite{ref:CDFbtojpsix},
as well as  inclusive muon production~\cite{ref:CDFInclMu} are all about a factor of two higher
than expected in theory.

The production of $\Upsilon$ at CDF~\cite{ref:CDFUpsi} is a factor of two to five
times larger than expected, and at ZEUS the photoproduction rate of  
$\Upsilon$ 
is a factor of  five  too high~\cite{ref:ZEUSBxsect}.

Various $b$ production mechanisms at a hadron collider
have very different kinematics.  At tree level
two $b$ quarks are produced with 
an azimuthal separation of 180 degrees.  In contrast to this a contribution from
gluon splitting results in both $b$ quarks recoiling 
against a gluon~\cite{ref:CDFBxsect}, giving a small azimuthal separation
of the $b$ and $\overline{b}$ quarks.
If these contributions are not computed correctly,
this can result in an improper determination of acceptance and hence the
cross sections can be extracted improperly.  

These contributions have been explored~\cite{ref:CDFBxsect} by
 measuring the
azimuthal separation of the muon decay from the $b$ 
quark and the jet from the $\bar{b}$ which was tagged using the impact 
parameter  of the tracks in the jet.
 The result shown in Figure~\ref{fig:bopencrossect} indicates
 that while the normalisation remains incorrect, the shape seems
to be correctly reproduced but with rather large errors in the sensitive
region. 

\subsection{Decays}
Among the measurements of rare decays, one of current interest is 
the CLEO~\cite{ref:cleodecay}
measurement of the branching ratio for $B_d^0 \rightarrow \pi^+\pi^-:
4.7^{+1.8}_{-1.5}\pm 0.6 \times 10^{-6}$. 
This is bad news for the measurement of $\sin{2\alpha}$ since
      estimates on the experimental 
errors assumed a branching ratio a factor of two larger. 
CDF~\cite{ref:CDFjpsipol,ref:CDFjpsipol2} 
has measured the angular distributions of the decay products
 in $B^0_d \rightarrow J/\psi K^{*0}$ 
to obtain the polarization of the $B^0_d$ and found that the P-odd component 
is small.
 This is good news
  for identifying channels with a well-defined CP eigenstate but the results
  are low in statistics.
CDF has also observed~\cite{ref:CDFBc1,ref:CDFBc2} 
 the $B_c$  and the lifetime measurement
indicates that $c$ quark decays first: $\tau_{Bc}=
  0.46^{+0.18}_{-0.16}\pm 0.03$ ps. Unfortunately, full reconstruction of
the $B_c$ in  
$B_c \rightarrow J/\psi \pi$ will be difficult because the lifetime
measurement implies a branching ratio that is ten times smaller.
 
\subsection{$B_d^0$ Mixing} 
A number of precise measurements of $B_d^0$ mixing at LEP, SLC and the
Tevatron exist.  The fraction of mixed mesons is given by
$F_{mix}(t)=\frac{1}{2}\left(1-\cos{\Delta m_d t}\right)$ and is measured as a
function of proper time.  
 In order
to measure mixing, the $B^0_d$ meson must be reconstructed or partially
reconstructed and its flavour must be tagged
at production and decay.  
Two major classes of tagging in $B$ mixing are 
same side and opposite side tagging.
Given a semileptonic decay on one side where the
lepton satisfied a trigger, one may
search for a semileptonic decay on the other side. Since the other lepton
is not biased by the trigger, it generally has lower momentum and is called a
slow lepton.  This tag is referred to as a Slow Lepton Tag (SLT).  The soft
fragmentation pion in the same hemisphere as the $B_d^0$ meson may be
identified and used to tag the initial $B_d^0$ state.
 This is called a same side tag (SST).  The charge
of particles in the opposite hemisphere may be summed with a momentum  or
lifetime weighting to provide a jet charge tag.
The CDF group has measurements that use
the following combinations of tags~\cite{ref:cdfbdmix}:
$D^*$ lep/SST;
lep/$Q^{jet}$, lep;
$e$/$\mu$; 
$\mu$/$\mu$; 
$D^*$ lep/lep;
$D^{\left(*\right)}$/lep.
 Results for these mixing measurements
are shown in Table~\ref{tab:cdfbmix}.  They agree with one
another and
with other measurements of $B$ mixing at LEP and SLD~\cite{ref:bmixwg}.
\begin{center}
\begin{table}[htb]
\caption{CDF $B_d^0$ mixing results}
\begin{indented}
\item[]\begin{tabular}{@{}ll}
\br
Tag & $ \Delta m_d$\\
\br
$D^*$ lep / SST                  & $ 0.471^{+0.078}_{-0.068}\pm0.034$ ps$^{-1}$ \\
lep /$Q^{jet}$, lep              & $ 0.500 \pm 0.052 \pm 0.043$ ps$^{-1}$ \\
$e$/$\mu$                        & $ 0.450 \pm 0.045 \pm 0.051$ ps$^{-1}$ \\
$\mu$/$\mu$                      & $ 0.503 \pm 0.064 \pm 0.071$ ps$^{-1}$ \\
$D^*$ lep /lep                   & $ 0.516 \pm 0.099 ^{+0.029}_{-0.035}$ ps$^{-1}$ \\
$D^{\left(*\right)}$ / lep   & $ 0.562 \pm 0.068 ^{+0.041}_{-0.050}$ ps$^{-1}$ \\
\br
Average                          & $0.495 \pm 0.026 \pm 0.025$  ps$^{-1}$ \\
\br
\end{tabular}
\end{indented}
\label{tab:cdfbmix}
\end{table}
\end{center}

\subsection{$B_s^0$ Mixing}
Similar tagging methods are used in the search for $B_s^0$ mixing.  This is
currently done by fitting the data with a fixed mixing frequency and allowing
the amplitude to vary as $A\cos{\Delta m_s t}$.  The result presented by the
LEP $B$ mixing working group is that
the world average $\Delta m_s> 14.3$ ps$^{-1}$ at 95\% CL.  Because a large
value of $\Delta m_s$ will result in different lifetimes for the two
eigenstates, the $B_s^0$ lifetime may be fitted for two components. Combined
results from  
LEP and CDF give $\Delta m_s <$ 40ps$^{-1}$ at 95\% CL.  Using the
world average $B_s^0$ lifetime, this gives $ 20.9 < x_s < 58.4.$
It is worth noting that CDF RunII with the Layer 00 upgrade can reach
a value of $x_s \simeq 60$.  Also, the SLD and LEP collaborations still have
analysis that take advantage of improvements in their tracking that could in
fact find a value of $\Delta m_s$ before the Tevatron takes data.

\subsection{CP Violation}
CDF~\cite{ref:cdf_sin2b} and OPAL~\cite{ref:opal_sin2b} have produced
 measurements of the CP asymmetry where a $B_d^0$
meson decays to the same eigenstate. This asymmetry is given by:
$$
A_{CP}\left(t\right)=\frac{\overline{B}^0_d\left(t\right)-B^0_d\left(t\right)}
{\overline{B}^0_d\left(t\right)+B^0_d\left(t\right)}=\sin{2\beta}\sin{\Delta m_d t}.
$$
In all cases, 
$$
A_{obs}=D A_{CP} = D \sin{2\beta}\sin{\Delta m_d t},
$$
where $D$ is the dilution due to mistagging and the error on the CP asymmetry
goes as:
$$\delta A_{CP} \sim \frac{1}{\sqrt{N \epsilon D^2}},$$
where $\epsilon  $ is the tagging efficiency or the fraction of events
tagged and $D= \left(R-W\right)/\left(R+W\right)$ is the dilution or the number
of right, $R$, minus wrong, $W$, tags.  
The analysis of CP asymmetry hinges on the
ability to tag accurately, maximising the dilution parameter. 

CDF reconstructs $J/\psi$ and $K^0_s$ with 
400 events observed.
Half of these are contained in the silicon tracker, the
SVX, so that there is a time-dependent measurement for 200 events.
Because the behaviour of $A_{CP}$ is a sine function,
and the largest background is at low lifetimes, most sensitivity to 
$\Delta\sin{2\beta}$ is in a lower background region at higher lifetimes.

In order to obtain $\sin{2\beta}$, the dilution must be measured. This is
done  using a  $B^+ \rightarrow J/\psi K^+$ sample to obtain the
 dilution in the jet charge  and slow lepton tags.
A lepton $D^*$ sample is used to determine the SST dilution.
A boost correction from Monte Carlo must
be added, resulting in a systematic uncertainty
$\Delta D = 0.013$.  The measured dilutions and contributions to the 
value of $\sin{2\beta}$ for each of the tagging methods is detailed in
  Table~\ref{tab:sin2b}. 

\begin{center}
\begin{table}[htb]
\caption{Breakdown of dilutions and $\sin{2\beta}$ by tag}
\begin{indented}
\item[]\begin{tabular}{@{}llll}
\br
Tag & $\epsilon$ & $D$ & $\sin{2\beta}$\\
\br
${\rm SST}_{\rm svx}$   & $35.5 \pm 3.7$ & $16.6 \pm  2.2 $ & $1.77^{+1.04}_{-1.01}$ \\
${\rm SST}_{\rm nosvx}$ & $38.1 \pm 3.9$ & $17.4 \pm  3.6 $ \\\mr
SLT                     & $ 5.6 \pm 1.8$ & $62.5 \pm 14.6 $ &$0.52^{+0.61}_{-0.75}$\\\mr
JetQ                    & $40.2 \pm 3.9$ & $23.5 \pm  6.9 $ &$-0.31^{+0.81}_{-0.85}$ \\\mr\mr
Global                  &\multicolumn{2}{c}{$\epsilon D^2=6.3\pm1.7$}&$0.79^{+0.41}_{-0.44}$\\
\br
\end{tabular}
\end{indented}
\label{tab:sin2b}
\end{table}
\end{center}

\begin{figure}[htb]
   \begin{center}
      \includegraphics[width=15cm]{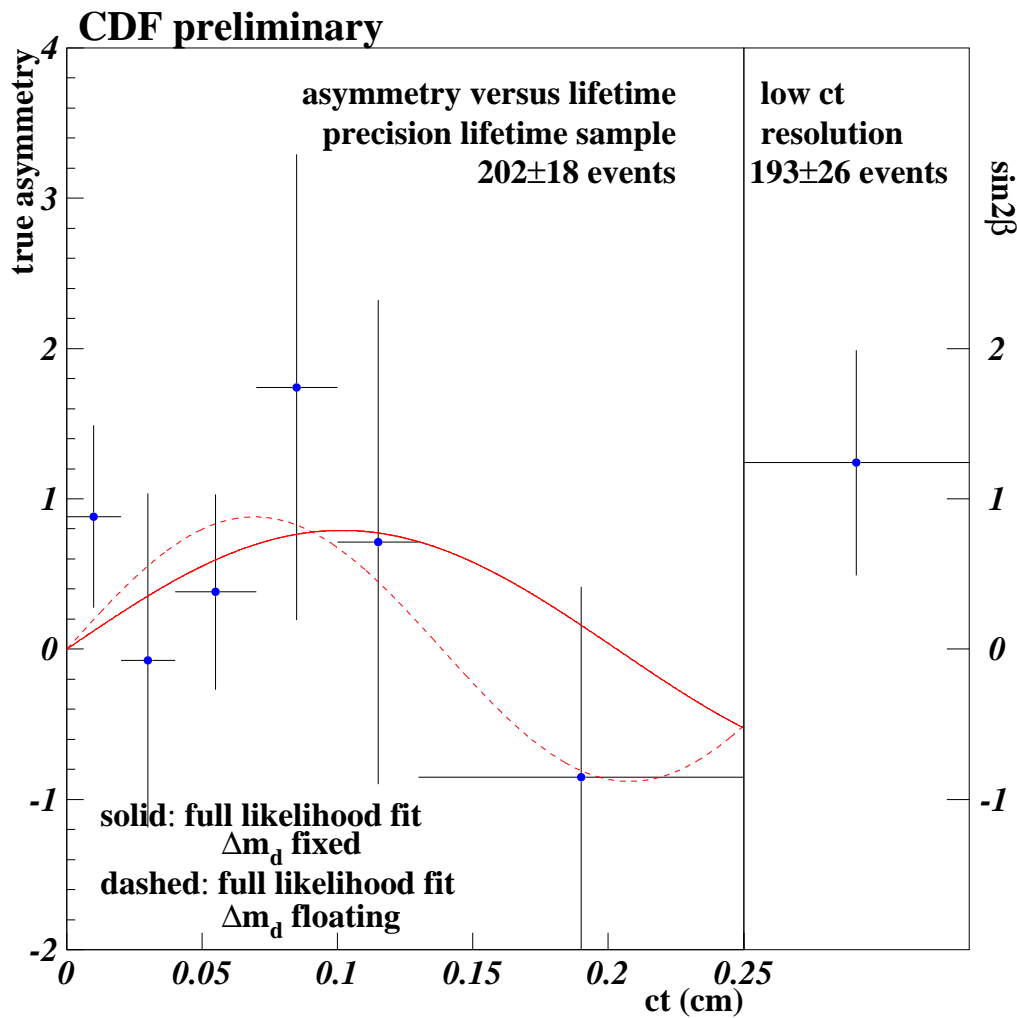}
   \end{center}
   \caption{CP asymmetry vs proper time as measured by CDF.}
   \label{fig:sin2b}
\end{figure}

The resulting CP asymmetry is plotted as a function of proper time in
Figure~\ref{fig:sin2b}.
The result is that $\sin{2\beta}=0.79\pm0.39\pm0.16$ 
or $0.0 < \sin{2\beta} < 1$ at 93\%CL.
This may be compared
to the OPAL measurement of
 $\sin{2\beta}=3.2^{+1.8}_{-2.0}\pm0.5$.
This is a proof of principle showing that in the hadron collider the error
is predominantly statistical.  Furthermore, since the
 systematic errors are statistics dominated, having more data will improve the
 measurement.

\section{Top at Hadron Colliders}

The decay of a top quark results in 
six jets, or four jets, a lepton and a neutrino or
two jets with two leptons and two neutrinos.  In each case, two of the jets are
$b$ jets.  These three cases comprise the major signatures for top that are
studied at the Tevatron.  Table~\ref{tab:TopSummary} shows the three modes,
the kind of signature used to distinguish each mode, the dominant source of
background, the branching ratios and the number of signal and background
events in each of the two detectors, CDF and D0.

\begin{center}
\begin{table}[htb]
\caption{Events in each detector for the various jet modes, signatures,
  backgrounds and branching fractions.}
\begin{indented}
\item[]\begin{tabular}{@{}llll}
\br
Mode              & Signature    & Bkgr                        & BR       \\
\mr
6 jet             & $b$,Kine     & QCD                         & 45\%     \\
            CDF   & 157          & 123                         &          \\
            D0    &  18          &   7                         &          \\
\mr
lep + $\geq$4 jet & $E_t^{miss}$ & $W +$ {\it jet}             &30\%     \\ 
                  & lep jets $b$ &                             &  \\
CDF               & 96           &  40                         & \\
        D0        & 30           &  12                         & \\
\mr 
lep + lep         & $E_t^{miss}$ & Drell Yan                   & 5\%\\
                  & lep lep 2jet & Z$\rightarrow \tau^+\tau^-$ &    \\
        CDF       & 13           &  4.4                        &   \\
       D0         &  5           &  1.4                        &   \\ 
\br
\end{tabular}
\end{indented}
\label{tab:TopSummary}
\end{table}
\end{center}

A number of measurements of the top quark have been performed.  These
include:

\begin{itemize}

\item Mass: $174.3 \pm 5.1$ GeV/c$^2$~\cite{ref:tmass};
\item $\sigma_{t\bar{t}}$: $5.9 \pm 1.7$ pb where theoretical predictions are
  4.7-5.5 pb~\cite{ref:txsect};
\item $V_{tb}$: $0.99 \pm 0.29 $~\cite{ref:tmass};
\item $W$ Helicity ($p_t^l$): $F=0.55 \pm 0.32 \pm 0.12$; where the theory
  predicts  0.7~\cite{ref:tmass};
\item Single Top: $\sigma\left(Wg \,\,\,{\rm fusion}\right) <$ 15.4 pb ,
  $\sigma\left(W^{+*}\rightarrow t \bar{b}\right) <$ 15.8 pb
   Theoretical predictions are
  1.7 and 0.7 pb respectively~\cite{ref:singtop}. 
   ALEPH~\cite{ref:alephopentop} 
    has reported 18 events on a background of
   10.  There is a  4.6\% chance that the background fluctuation can give
   these observed events and they are looking forward to collecting more
   statistics.  The Standard Model branching ratio is $10^{-6}$.
\item Extensive kinematic studies~\cite{ref:topkine} have shown agreement with 
      the expectation. Values studied are leading jet $E_t$, jet rapidity
      separations, etc.
\item The invariant mass~\cite{ref:singtop} of the top quark pair, $m_{t\bar{t}}$, has been
      studied  and a limit on a $Z^{\prime}$ of $m_{Z'}>650$ GeV/c$^2$
      has been set.  More generally, the limits from the data on 
      $\sigma_X \times {\rm B}\left(X\rightarrow t\bar{t}\right)$ for
       various hypothetical masses $m_X$ of a new particle that decays to $t\bar{t}$
       are presented.
\item A limit on branching ratios in rare top decays~\cite{ref:tmass} has been set with
B($t\rightarrow q\gamma$)$ < 0.032$ and
B($t\rightarrow qZ$)$ < 0.33 $.

\end{itemize}

The value of the top cross section at the Tevatron is slightly high compared 
to theory but
within statistical errors.  However,
each of the published cross section
values for the individual modes for each experiment
shows that all but one of them is high.
Recent re-analysis of the data shows that the cross section values
are tending to come down to the expected values.

\section{Conclusions}

In surveying the present state of heavy flavours, one is first struck by
the fact that the heavy quark cross sections  are all too high.
The top cross section agrees with theory but the errors are large.
The $c$ quarks are used at HERA to successfully set the scale for 
QCD and measure the gluon structure function; however, a consistent
explanation of the production cross sections 
at HERA and the Tevatron is not possible.

The $B$  meson and baryon lifetime ratios show agreement with 
theory and the spectator picture. A measurement of $B_s$ mixing is coming soon
with current estimates of $ 20.9 < x_s < 58.4 $.  SLD and LEP may use
optimized tracking algorithms and a unified analysis to find the value of 
$\Delta m_s$ before the Tevatron, where it will most certainly be measured
with precision.
CP violation measurements in  hadron machines are 
 not limited by systematics and the $e^+e^-$ machines are 
  starting up.  With the data that will be collected at the Tevatron, HERA,
  and at the $e^+ e^-$ machines (BELLE, PEPII, and CESR)
 there will be a fantastic increase in information on heavy flavours in the
 next few years.

\ack
I wish to thank the organizers for putting together a very
good and enjoyable workshop.



\end{document}